# An on-chip tunable micro-disk laser fabricated on Er$^{3+}$ doped lithium niobate on insulator (LNOI)

ZHE WANG,[1,3,4] ZHIWEI FANG,[2,8] ZHAOXIANG LIU,[2,9] WEI CHU,[2] YUAN ZHOU,[1,3] JIANHAO ZHANG,[1,3] RONGBO WU,[1,3] MIN WANG,[2] TAO LU,[5] AND YA CHENG[1,2,6,7,10]

[1]State Key Laboratory of High Field Laser Physics and CAS Center for Excellence in Ultra-intense Laser Science, Shanghai Institute of Optics and Fine Mechanics (SIOM), Chinese Academy of Sciences (CAS), Shanghai 201800, China
[2]XXL—The Extreme Optoelectromechanics Laboratory, School of Physics and Materials Science, East China Normal University, Shanghai 200241, China
[3]Center of Materials Science and Optoelectronics Engineering, University of Chinese Academy of Sciences, Beijing 100049, China
[4]School of Physical Science and Technology, ShanghaiTech University, Shanghai 200031, China
[5]Department of Electrical and Computer Engineering, University of Victoria, Victoria, BC, V8P 5C2, Canada
[6]Collaborative Innovation Center of Extreme Optics, Shanxi University, Taiyuan 030006, China.
[7]Collaborative Innovation Center of Light Manipulations and Applications, Shandong Normal University, Jinan 250358, People's Republic of China
[8]zwfang@phy.ecnu.edu.cn
[9]zxliu@phy.ecnu.edu.cn
[10]ya.cheng@siom.ac.cn

We demonstrate a C-band wavelength-tunable microlaser with an Er$^{3+}$ doped high quality (~$1.02 \times 10^6$) lithium niobate microdisk resonator. With a 976 nm continuous-wave pump laser, lasing action can be observed at a pump power threshold as low as ~250 µW at room temperature. Furthermore, the microdisk laser wavelength can be tuned by varying the pump laser power, showing a tuning efficiency of ~-17.03 pm/mW at low pump power blow 13 mW, and 10.58 pm/mW at high pump power above 13 mW.

## 1. INTRODUCTION

Featured by its broad optical transparency window (0.35-5 µm), high nonlinear coefficient ($d_{33} = -41.7 \pm 7.8 \, pm/V@\lambda = 1.058$ µm), high refractive index (~2.2), and large electro-optical effect ($r_{33} = 30.9 pm/V@\lambda = 632.8 nm$), lithium niobate on insulator (LNOI) is a promising material platform for photonic integrated circuit (PIC) [1-3]. So far, a variety of LNOI-based PIC devices have been demonstrated including electro-optic modulators, nonlinear optical devices, and quantum optical devices which have made strong impacts on a wide range of applications ranging from optical communications, microwave photonics and quantum optics to optical computing and metrology [4-9]. All of these LNOI PIC devices are demonstrated with external light sources. It is highly desirable to realize on-chip generation and integration of light sources into the LNOI PIC devices which enables more compact, more efficient, and more stable PIC devices. For this reason, it is of vital importance to realize on-chip microlasers with a wavelength tunability on the LNOI substrate. It should be noted that so far C-band microlasers based on microresonator have been successfully fabricated in indium phosphide (InP), indium arsenide (InAs), Er$^{3+}$-doped silica by photolithography methods [10-18]. Unlike these materials, high quality (Q) LNOI-based microresonators have only been demonstrated recently [19-23]. Here, we demonstrate a C-band microlaser on an Er$^{3+}$-doped LNOI chip. The diameter of the LN microdisk is 200 µm. The microdisk resonator is pumped at 976 nm through a tapered optical fiber and the lasing threshold is 250 µW at room temperature. The low threshold is result of the high cavity optical Q (~$1.02 \times 10^6$), which leads to a large intracavity pump intensity. We also observed a laser wavelength tuning efficiency of ~-17.03 pm/mW at low pump power blow 13 mW, and 10.58 pm/mW at high pump power above 13 mW.

## 2. DEVICE FABRICATION

In our experiment, the on-chip $Er^{3+}$ doped LN microdisk was fabricated on an $Er^{3+}$ doped Z-cut LN thin film wafer with a thickness of 600 nm. A 3-inch $Er^{3+}$ doped LN wafer (concentration of $Er^{3+}$ ions ~1 mol%) which is purchased from Shanghai Daheng Optics and Fine Mechanics Co., Ltd was ion sliced into a LN thin film with a thickness of 600 nm by Jinan NANOLN Co., Ltd. The $Er^{3+}$-doped LN thin film was bonded onto a silica layer with a thickness of ~2 μm, and the silica layer was grown on a 0.5 mm-thick undoped crystalline LN substrate [24]. As shown in Fig. 1(a), although the undoped LN wafer appears almost colorless, the $Er^{3+}$ doped LN wafer appears reddish. The configuration of $Er^{3+}$-doped LN-on-insulator ($Er^{3+}$ LNOI) was shown in Fig. 1(b), on top of which a 600-nm-thickness layer of chromium (Cr) film was deposited by magnetron sputtering method. The fabrication process includes five steps, as illustrated in Fig. 1(c)-(f). First, the Cr film on the $Er^{3+}$LNOI sample was patterned into a disk-shaped mask using space-selective femtosecond laser (PHAROS, LIGHT CONVERSION Inc.). Subsequently, the chemo-mechanical polishing (CMP) process was performed to fabricate the LN microdisk by a wafer lapping polishing machines (NUIPOL802, Kejing Inc.). In this step, the LN film underneath the Cr mask can be preserved whereas the LN in the opening area is completely removed. The CMP process allows to create a LN microdisk with extremely smooth rim for high Q factors. Then, the Cr thin film was removed with a chemical wet etching process and a secondary CMP process was performed for thinning the LN disk as well as cleaning the ablation debris on the disk surface. Finally, the fabricated structure was immersed in a buffered hydrofluoric acid (HF) solution to partially under etch the silica layer into the shape of a pillar supporting the LN microdisk. It takes about 2 hours in total to produce the device. More details about the LN microdisk fabrication can be find in Ref. 21-23.

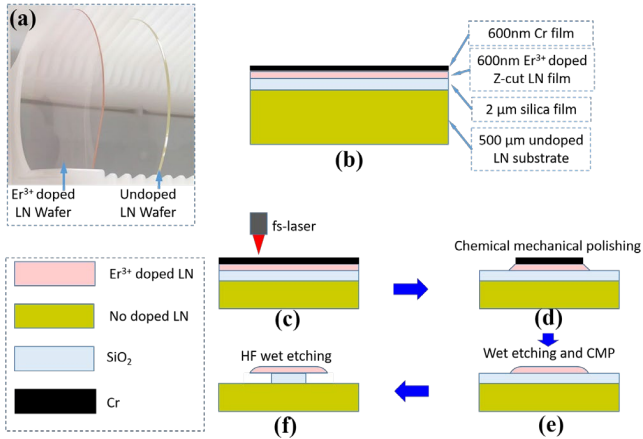

Fig. 1. Schematic of the $Er^{3+}$ doped LN microdisk fabrication. (a) $Er^{3+}$ doped LN wafer and undoped LN wafer. (b) The Cr film was deposited on the surface of the $Er^{3+}$-doped lithium niobate on insulator. (c)-(d) Patterning the Cr thin film into a circular mask using femtosecond laser microfabrication, and transferring the circular disk-shaped mask to $Er^{3+}$ LNOI by CMP. (e) Removing the Cr thin film with chemical wet etching process, and reducing the thickness of LN disk by a secondary CMP. (f) Chemical wet etching of the sample to undercut the $SiO_2$ beneath the LN microdisk, forming the freestanding $Er^{3+}$ LN microdisk resonator supported by the $SiO_2$ pedestal.

## 3. DEVICE CHARACTERIZATION

To characterize the laser performance of the $Er^{3+}$-doped LN microdisk, we used an experimental setup as shown in Fig. 2(a). Here, a continuous-wave C-band tunable laser (TLB 6728, New Focus Inc.) was used for characterizing the Q factor of the microdisk. Alternatively, a diode laser (CM97-1000-76PM, Wuhan Freelink Opto-electronics Co., Ltd.) operated at the wavelength ~976nm was chosen to pump the $Er^{3+}$-doped LN microdisk. The polarization state of the pump laser was adjusted by an in-line fiber polarization controller. A tapered fiber with a waist of 1 μm was used to couple the light into and out of the fabricated $Er^{3+}$-doped LN microdisk. The output beam was directed to an optical spectrum analyzer (OSA: AQ6370D, YOKOGAWA Inc.) for optical spectral analysis. A power meter (PM100D, Thorlabs Inc.) was used for input pump power monitoring. The Q factor of an resonant mode was measured by a photodetector (New focus 1811-FC-AC, Newport Inc.). The LN microdisk features a wedge which a slide wall angle of 10°, as evidenced by the inset of Fig. 2(a). Fig. 2(b) illustrates the measured transmission spectrum near the wavelength of 1553.16 nm at a low pump power where the thermal broadening effect is negligible. The Q factor was determined to be $1.02 \times 10^6$ through a Lorentzian fitting. Expected, the excessive absorption from the $Er^{3+}$ ions reduces the optical Q compared to the undoped microdisk [21].

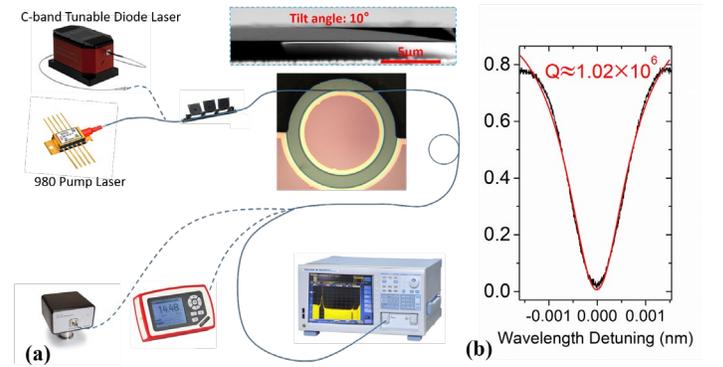

Fig. 2. Experimental testing setup: (a) The tunable diode laser or the 976nm pump laser are coupled into and out of the $Er^{3+}$-doped microdisk via a tapered optical fiber, Inset: Side view SEM of the disk with a wedge angle of 10°. (b) The Lorentzian fitting indicating the Q-factors of $1.02 \times 10^6$ of the microdisk as measured at 1553.16 nm wavelength.

The energy level diagram of the $Er^{3+}$ ions as well as the absorption and the emission transition is depicted in Fig. 3(a). It has been established that the intermediate state $^4I_{11/2}$ and the $^4F_{7/2}$ state can be accessed through resonantly excitation and excited state absorption with the pump laser at 976 nm, respectively. The $^2H_{11/2}$, $^4S_{3/2}$, and $^4I_{13/2}$ states can be populated by the effective nonradiative relaxations of the $^4F_{7/2}$, $^2H_{11/2}$, and $^4I_{11/2}$. The fluorescence emissions at ~530 nm, ~550 nm in Fig. 3(b) and that at ~1550 nm in Fig. 3(c) can be attributed to the transitions of $^2H_{11/2} \rightarrow ^4I_{15/2}$, $^4S_{3/2} \rightarrow ^4I_{15/2}$ and $^4I_{13/2} \rightarrow ^4I_{15/2}$, respectively [25]. As shown in Fig. 3(b) and (c) with very slight pump power at the wavelength of 976 nm. The abundant fluorescence lines indicate that the ion concentration of the $Er^{3+}$-doped LN microdisk is sufficiently high for generating lasers if strong pump laser is applied.

Fig. 4(a) shows a typical pump and laser emission spectrum collected in the 800-1670 nm spectral range. Clearly, the wavelength of the pump is ~976 nm and the laser emission is observed at ~1560 nm. The inset in Fig. 4(a) displays the strong green upconversion fluorescence of the microdisk, which was taken using a CMOS camera (DCC3240C, Thorlabs Inc.) mounted onto an optical microscope. The optical beam path of the pump laser in the microdisk can clearly be seen in the inset of Fig. 4(a). Fig. 4(b) is the enlarged spectrum near 1560 nm, featuring several laser lines at the C-band telecommunication wavelengths. The lasing threshold can be determined by integrating the intensity of the emission lines over the 1560-1567 nm spectral range and plotting the integrated laser intensity as a function of the pump laser which is dropped into the cavity, as shown in Fig. 4 (c). The lasing threshold is observed around 250 μW, and a linear behavior is observed above the threshold.

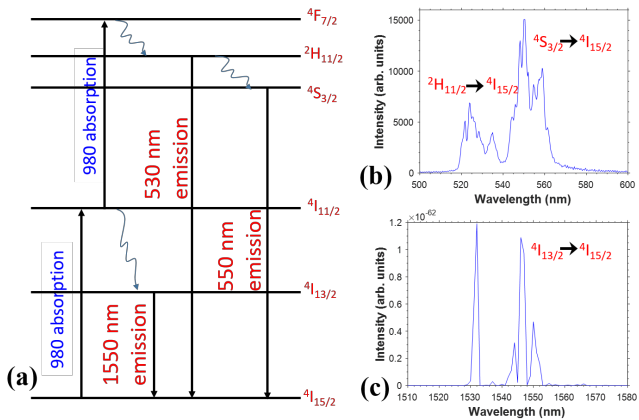

Fig. 3 Fluorescence spectra under 976 nm pump. (a) Schematic of the energy level diagram of the Er[3+] ions as well as the absorption and the emission transition. (b) Visible upconversion and (c) infrared downconversion emission spectra of the Er[3+]-doped LN microdisk.

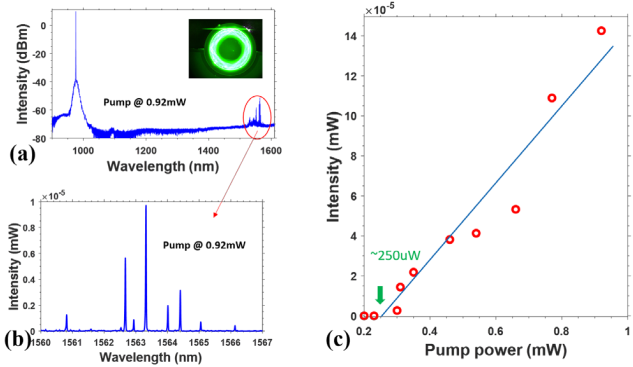

Fig.4. The lasing threshold characteristics of the Er[3+]-doped LN microdisk. (a)The spectrum of the Er[3+] doped laser with the pump power at 0.92mW, the inset displays the strong green upconversion fluorescence of the microdisk. (b)The enlarge spectrum around 1560 nm shows the multimode lasing spectrum. (c)The dependence of the intensity of all emission lines on the absorbed pump power at 976 nm, where the experimental data is shown as red circles, and the blue curve is a linear fitting, the lasing threshold is 250 μW.

The lasing mode of the Er[3+]-doped LN microdisk shows a strong dependence on the pump laser power. As shown in Fig. 5(a), at the pump laser power of 23.1mW, the laser behaves like a single frequency lasing emission at the wavelength around 1551 nm accompanied with several weak satellite emission lines. This should be a result of the strong competition between the lasing modes of different gain efficiencies.

It is well known that the high optical intensity inside the LN microdisk, can lead to both significant photorefractive and thermo-optic effects [26-28], which provides us an opportunity to efficiently tune the laser wavelength merely by changing the pump laser power. As shown in Fig. 5(b), the quasi-single frequency lasing wavelength can be tuned smoothly with the increasing pump power. When the input pump power increases from 4.05 mW to 9.8 mW, blue-shift of laser wavelength is observed, indicating that the photorefraction effect plays a leading role for resonance shift. The photorefraction effect manifests itself with a decrease of the refractive index depending on the optical intensity inside the microdisk. When the input power increases further from 17.7 mW to 40.6 mW, red-shift of the laser wavelength is observed. This implies that the thermo-optic effect plays the leading role at high pump intensity. To determine the tuning efficiency, the lasing spectra of the Er[3+]-doped LN microdisk were recorded at various input powers as shown in Fig. 5(c). A linear dependence of the emission wavelength on the pump power is observed, showing a tuning efficiency of -17.03 pm/mW for low pump powers blow 13 mW (corresponding to -2.1231GHz/mW), and 10.58 pm/mW for high pump power above 13 mW (corresponding to 1.319 GHz/mW). The observation indicates that the Er[3+]-doped LN microdisk laser provides an efficient and convenient method for all optical tuning of the on-chip laser wavelength.

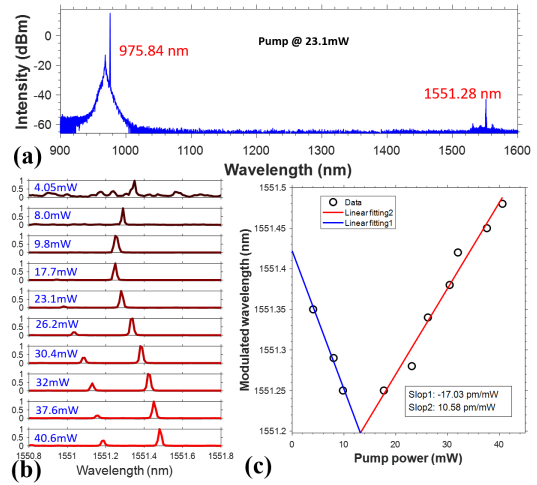

Fig. 5 Power tuning characteristics of the Er[3+] doped LN microdisk. (a)The spectrum of the Er[3+]-doped laser with the pump power at 82.3mW. (b)Recorded lasing spectra of the microdisk with the increasing input powers. (c)The modulated wavelength of the microdisk laser as function of the pump power, where the experimental data is shown as black circles, the blue and red curve is a linear fitting, with a slope of -17.03 pm/mW for low pump powers blow 13 mW and 10.58 pm/mW for high pump power above 13 mW (corresponding to -2.1231GHz/mW and 1.319 GHz/mW).

## 4. CONCLUSION

In summary, we demonstrate an Er[3+] doped LN microdisk C-band laser on a lithium niobate thin film chip. A tuning efficiency of ~-17.03 pm/mW is observed at low pump power blow 13 mW, whereas the tuning efficiency dramatically changes to ~10.58 pm/mW at high pump power above 13 mW. The LN microdisk laser can be integrated with other nanophotonics structures to construct LNOI-based PIC devices to serve as an on-chip tunable laser source.

**Funding**.
National Key R&D Program of China (2019YFA0705000, 2018YFB2200400), National Natural Science Foundation of China (Grant Nos. 11822410, 11874154, 11874375, 11734009, 61761136006, 11674340, 61675220, 61590934,), the Strategic Priority Research Program of Chinese Academy of Sciences (Grant No. XDB16030300), the Key Project of the Shanghai Science and Technology Committee (Grant Nos. 18DZ1112700, 17JC1400400), Shanghai Municipal Science and Technology Major Project (Grant No.2019SHZDZX01), the Shanghai Pujiang Program (Grant No. 18PJ1403300), the Shanghai Rising-Star Program (Grant No. 17QA1404600), the Key Research Program of Frontier Sciences, Chinese Academy of Sciences (Grant No. QYZDJ-SSW-SLH010), and the Fundamental Research Funds for the Central Universities

(2018FZA5004), Nature Science and Engineering Research Council of Canada (NSERC) Discovery (Grant No. RGPIN-2020-05938).